\documentclass[aps,prc,twocolumn,superscriptaddress]{revtex4}

\usepackage[dvips]{graphicx}

\def\EqTextStyle{\hspace{.25in}\parbox[h]{2.875in}}
\def\beginTable{\begin{table}[ht]\begin{center}}
\def\endTable{\end{center}\end{table}}
\def\beginFigure{\begin{figure}[ht]\begin{center}}
\def\endFigure{\end{center}\end{figure}}

\def\dedx {$dE/dx$}
\def\pt   {$p_t$}
\def\y    {$y$}

\def\cm      {~cm}
\def\gev     {~GeV}
\def\gevc    {~GeV/c}
\def\gevcc   {~GeV/c$^2$}
\def\mev     {~MeV}

\def\mevcc   {~MeV/c$^2$}
\def\evcc    {~eV/c$^2$}

\def\piz     {$\pi^0$}
\def\gzepz   {${\gamma}Z{\rightarrow}e^-e^+Z$}
\def\pizgg   {$\pi^0\rightarrow\gamma\gamma$}

\def\sqnn    {$\sqrt{s_{_\mathrm{NN}}}$}
\def\au      {$^{^{197}}$Au}
\def\auau    {\au+\au}
\def\pb      {$^{^{208}}$Pb}
\def\pbpb    {\pb+\pb}
\def\armGas  {10\% CH$_4$ and 90\% Ar}

\begin{document}

\title{Photon and neutral pion production in Au+Au collisions at \sqnn~=~130\gev}

\affiliation{Argonne National Laboratory, Argonne, Illinois 60439}
\affiliation{Brookhaven National Laboratory, Upton, New York 11973}
\affiliation{University of Birmingham, Birmingham, United Kingdom}
\affiliation{University of California, Berkeley, California 94720}
\affiliation{University of California, Davis, California 95616}
\affiliation{University of California, Los Angeles, California 90095}
\affiliation{Carnegie Mellon University, Pittsburgh, Pennsylvania 15213}
\affiliation{Creighton University, Omaha, Nebraska 68178}
\affiliation{Nuclear Physics Institute AS CR, \v{R}e\v{z}/Prague, Czech Republic}
\affiliation{Laboratory for High Energy (JINR), Dubna, Russia}
\affiliation{Particle Physics Laboratory (JINR), Dubna, Russia}
\affiliation{University of Frankfurt, Frankfurt, Germany}
\affiliation{Indiana University, Bloomington, Indiana 47408}
\affiliation{Insitute  of Physics, Bhubaneswar 751005, India}
\affiliation{Institut de Recherches Subatomiques, Strasbourg, France}
\affiliation{University of Jammu, Jammu 180001, India}
\affiliation{Kent State University, Kent, Ohio 44242}
\affiliation{Lawrence Berkeley National Laboratory, Berkeley, California 94720}\affiliation{Max-Planck-Institut f\"ur Physik, Munich, Germany}
\affiliation{Michigan State University, East Lansing, Michigan 48824}
\affiliation{Moscow Engineering Physics Institute, Moscow Russia}
\affiliation{City College of New York, New York City, New York 10031}
\affiliation{NIKHEF, Amsterdam, The Netherlands}
\affiliation{Ohio State University, Columbus, Ohio 43210}
\affiliation{Panjab University, Chandigarh 160014, India}
\affiliation{Pennsylvania State University, University Park, Pennsylvania 16802}
\affiliation{Institute of High Energy Physics, Protvino, Russia}
\affiliation{Purdue University, West Lafayette, Indiana 47907}
\affiliation{University of Rajasthan, Jaipur 302004, India}
\affiliation{Rice University, Houston, Texas 77251}
\affiliation{Universidade de Sao Paulo, Sao Paulo, Brazil}
\affiliation{University of Science \& Technology of China, Anhui 230027, China}
\affiliation{Shanghai Institute of Nuclear Research, Shanghai 201800, P.R. China}
\affiliation{SUBATECH, Nantes, France}
\affiliation{Texas A\&M, College Station, Texas 77843}
\affiliation{University of Texas, Austin, Texas 78712}
\affiliation{Valparaiso University, Valparaiso, Indiana 46383}
\affiliation{Variable Energy Cyclotron Centre, Kolkata 700064, India}
\affiliation{Warsaw University of Technology, Warsaw, Poland}
\affiliation{University of Washington, Seattle, Washington 98195}
\affiliation{Wayne State University, Detroit, Michigan 48201}
\affiliation{Institute of Particle Physics, CCNU (HZNU), Wuhan, 430079 China}
\affiliation{Yale University, New Haven, Connecticut 06520}
\affiliation{University of Zagreb, Zagreb, HR-10002, Croatia}
\author{J.~Adams}\affiliation{University of Birmingham, Birmingham, United Kingdom}
\author{C.~Adler}\affiliation{University of Frankfurt, Frankfurt, Germany}
\author{M.M.~Aggarwal}\affiliation{Panjab University, Chandigarh 160014, India}
\author{Z.~Ahammed}\affiliation{Purdue University, West Lafayette, Indiana 47907}
\author{J.~Amonett}\affiliation{Kent State University, Kent, Ohio 44242}
\author{B.D.~Anderson}\affiliation{Kent State University, Kent, Ohio 44242}
\author{M.~Anderson}\affiliation{University of California, Davis, California 95616}
\author{D.~Arkhipkin}\affiliation{Particle Physics Laboratory (JINR), Dubna, Russia}
\author{G.S.~Averichev}\affiliation{Laboratory for High Energy (JINR), Dubna, Russia}
\author{S.K.~Badyal}\affiliation{University of Jammu, Jammu 180001, India}
% ...

\collaboration{STAR Collaboration}\homepage{www.star.bnl.gov}\noaffiliation



\begin{abstract}

We report the first inclusive photon measurements about mid--rapidity ($|$\y$|$$<$0.5) from {\auau} collisions at {\sqnn} = 130{\gev} at RHIC. Photon pair conversions were reconstructed from electron and positron tracks measured with the Time Projection Chamber (TPC) of the STAR experiment. With this method, an energy resolution of $\Delta$E/E $\approx$ 2\% at 0.5{\gev} has been achieved. Reconstructed photons have also been used to measure the transverse momentum (\pt) spectra of {\piz} mesons about mid--rapidity ($|$\y$|$$<$1) via the {\pizgg} decay channel. The fractional contribution of the {\pizgg} decay to the inclusive photon spectrum decreases by 20\% $\pm$ 5\% between {\pt} = 1.65{\gevc} and {\pt} = 2.4{\gevc} in the most central events, indicating that relative to {\pizgg} decay the contribution of other photon sources is substantially increasing.

\end{abstract}



\maketitle

Relativistic heavy ion collisions provide the opportunity to excite matter into extreme conditions in the laboratory. Of the particles which emerge from these collisions, photons are considered to be one of the most valuable probes of the dynamics and properties of the resulting systems \cite{shuryak1,thesis,brem_space_time,WA98_direct_photons,WA98_long,directPhotonRecentReview}. Unlike hadrons, which have large interaction cross sections in dense matter, photons only interact electromagnetically and consequently have a long mean free path. This path length is typically much larger than the transverse size of the matter created in nuclear collisions \cite{hadronGas}. Therefore, with high probability, photons will escape from the system undisturbed, retaining information about the physical conditions under which they were created.

Photons are produced in all stages of heavy ion collisions \cite{shuryak1,brem_space_time,hot_glue,flash,hadronGas,thermal_photons_with_transverse_flow,a1_meson,two_loop,two_loop2,phot_inter,directPhotonRecentReview}, from the first instant when the quarks and gluons of the opposing nuclei interact, through to long lived electromagnetic decays of final state hadrons. The production rate of photons during various stages of the created system has been theoretically calculated for a variety of initial conditions and scenarios. It has been demonstrated that the emission rate from a hadron gas can be comparable to that expected from quark--gluon Compton and quark-antiquark annihilation processes in a net-baryon free system of deconfined quarks and gluons \cite{hadronGas}. However, recent two--loop calculations which include the quark bremsstrahlung process predict that photon production rates in quark matter exceed those indicated by former one--loop calculations that only account for the Compton and annihilation processes \cite{two_loop,two_loop2}. These calculations indicate ``that the emissions from quark matter can outshine those from the hadronic matter'' \cite{two_loop2}, and that near and above {\pt} = 2{\gevc} the contribution from hard scattered partons becomes more abundant than thermal photons from a hot hadronic gas at RHIC energy \cite{phot_inter}. Theoretical calculations and predictions like these underscore the importance of measuring photon spectra across a wide range of {\pt} to investigate the matter created in heavy ion collisions. The measurements presented in this paper extend above and below the interesting region around {\pt} = 2{\gevc}.

Of all photon production mechanisms, late stage electromagnetic decays of hadrons are the dominant source. At CERN SPS energies, photons from {\piz} and $\eta$ decays account for $\sim$97\% \cite{WA98_direct_photons} of the inclusive photon spectrum. The remaining 3\% arises from a combination of other sources, including electromagnetic decays of other hadrons such as the $\omega$, $\eta'$ and $\Sigma^0$. For these particles, thermal models that describe hadron production must be used to estimate the yields since their production rates have not yet been measured in heavy ion collisions at these energies. Measurement of the $\eta'\rightarrow\rho\gamma$ and $\Sigma^0\rightarrow\Lambda\gamma$ decays appears to be promising with the energy resolution afforded by the photon reconstruction technique described in this paper. By estimating or measuring the yields of such particles, their contribution to the single photon spectrum from electromagnetic decays can be calculated. However, the precision necessary to disentangle the rate of direct photon production from the rate of photons produced by electromagnetic decays is an experimental and theoretical challenge.

STAR has begun to address this challenge by measuring the spectra of both photons and {\piz}s. Photons were measured by reconstructing pair conversions, \gzepz, with the electron and positron daughters detected in the STAR TPC. Reconstructed photons were used in turn to measure the rate of {\piz} production via the {\pizgg} decay channel. This paper discusses the techniques employed and the resulting spectra of photons and {\piz}s that were measured. A full discussion of all cuts and variables used in this analysis may be found in \cite{thesis}.

\section{Data Analysis}
The data presented in this paper were recorded by the STAR collaboration during the first \sqnn~=~130{\gev} Au+Au run at RHIC. Events that had a primary collision vertex position less than 100{\cm} and 150{\cm} distant from the geometric center of the TPC along the beam axis ($z$--axis) were selected for the photon and {\piz} analyses respectively. Details of the STAR geometry are presented in Fig. \ref{fig::converters} and discussed in \cite{star_geometry,star_geometry2}. For the event sample used for {\piz} measurements, which were limited by statistical uncertainties, approximately 87\% of the events in the minimum bias (least trigger--biased) data set passed the $z$ vertex requirement. With this range of collision vertices, part of the support structure for the Silicon Vertex Tracker (SVT) could be utilized as a converter, as shown in Fig. \ref{fig::converters}. In addition, the inner field cage and gas (\armGas) of the STAR TPC were also used as converters. The combined material from both detectors resulted in an average conversion probability of approximately 1\% during the data run of year 2000. Although this conversion probability was low, it was compensated by the complete 2$\pi$ azimuthal acceptance of the STAR TPC.

\beginFigure
\begin{minipage}[ht]{7.7cm}
\begin{center}
\includegraphics*[width=.9\textwidth]{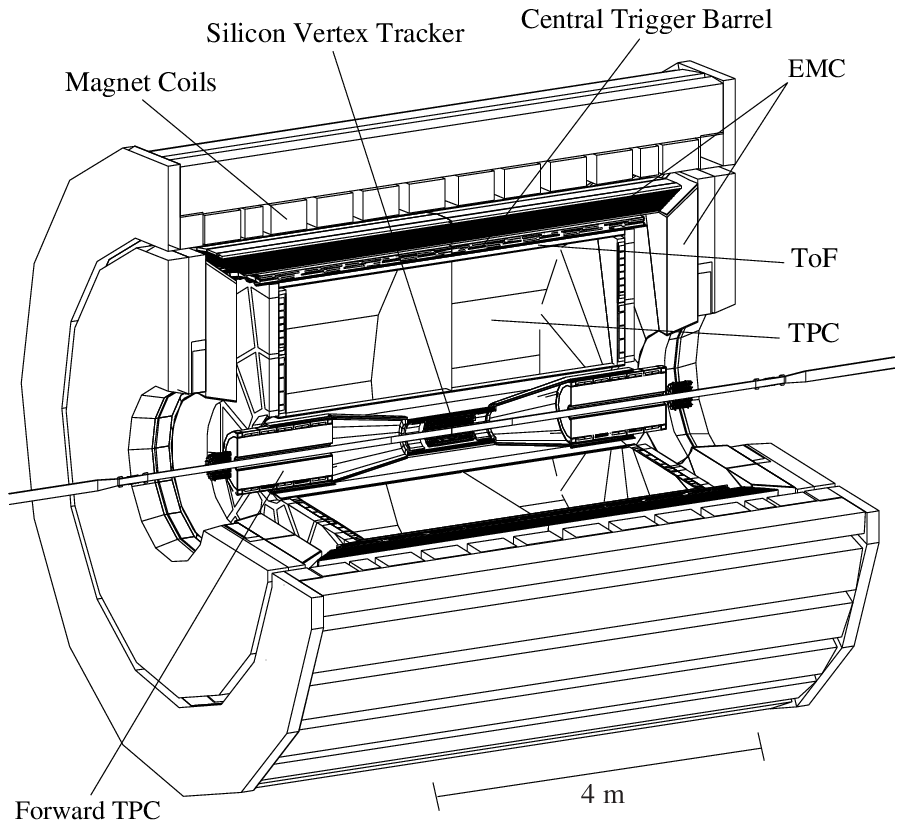}
\includegraphics*[width=\textwidth]{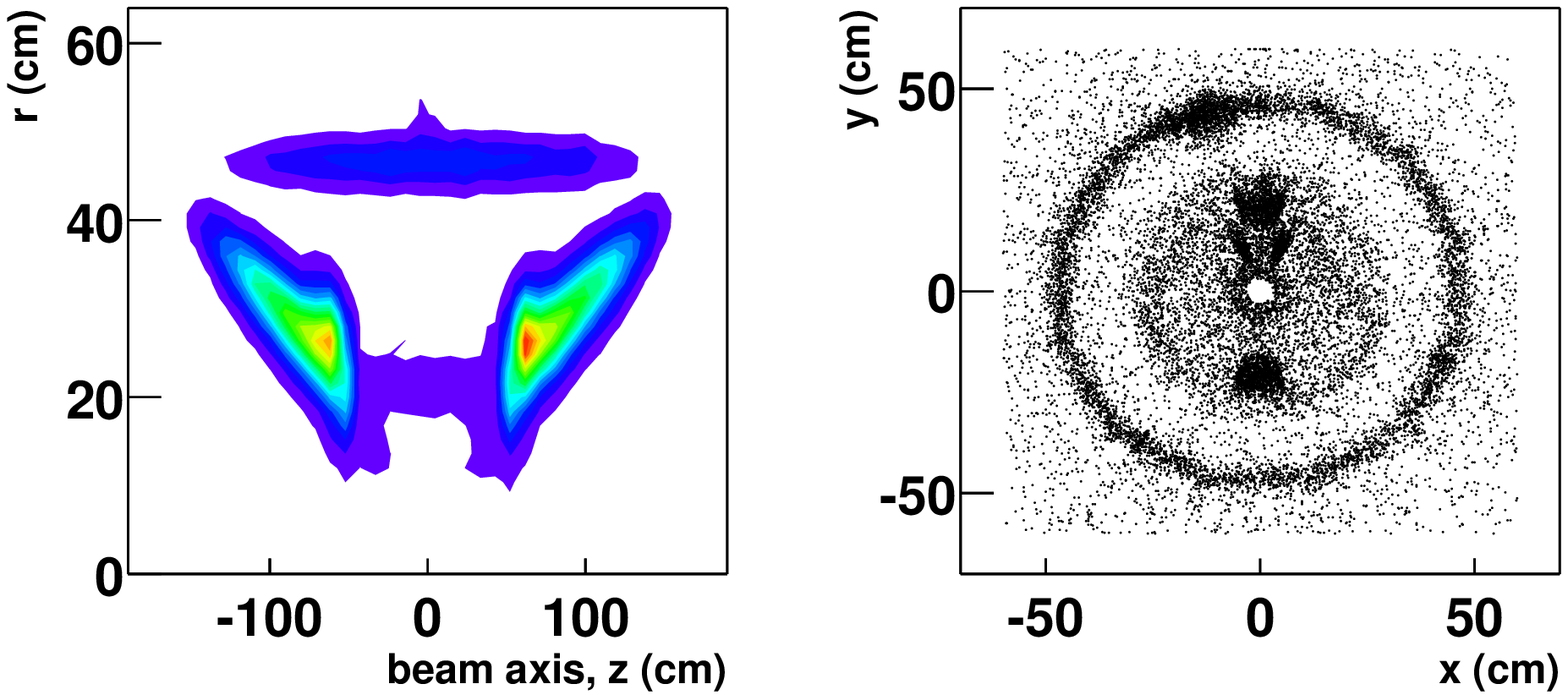}
\end{center}
\end{minipage}
\caption{Top figure: layout of the STAR experiment. Lower figures: density profiles of photon conversion points show the layout of the detector material. The structure at $r$ = 46.5{\cm} is the inner field cage of the cylindrical TPC, while below $r$ = 40{\cm} the SVT support cones and material are apparent.}
\label{fig::converters}
\endFigure

As discussed in \cite{flow_paper}, the definition of collision centrality was based on the number of reconstructed primary tracks in the pseudo--rapidity range $|$$\eta$$|$$<$0.75. Using this as a basis, four centrality classes were defined, common to both the photon and {\piz} analyses. They were an inclusive minimum bias (0--85\% of the total inelastic hadronic cross section), peripheral (34\%--85\%), mid central (34\%--11\%) and central (0--11\%). These centrality classes were selected to allow the extraction of  {\piz} yield over a wide range of {\pt} (0.25$<$\pt$<$2.5\gevc) in independent regions of centrality. They contained 328980, 198196, 87484 and 449095 events, respectively.

\beginFigure
\includegraphics*[width=.367\textwidth]{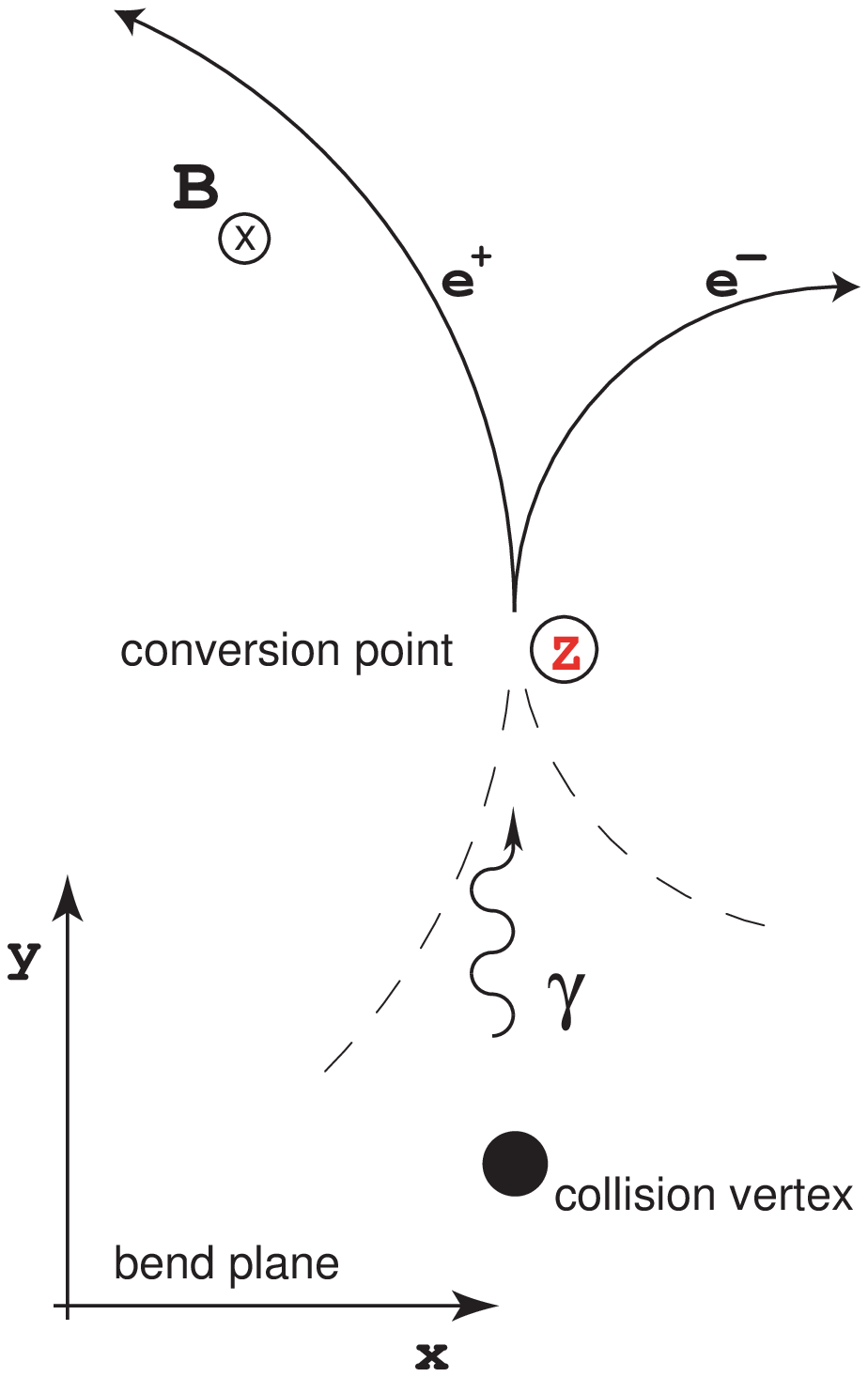}
\caption{Schematic diagram of \gzepz}
\label{fig::phot_con_schematic}
\endFigure

\subsection{Reconstructing Photon Pair Conversions \gzepz \label{sec::reconPhot}}
The dominant interaction process for photons with a total energy above 10{\mev} is pair conversion, {\gzepz} (Fig. \ref{fig::phot_con_schematic}). Pair conversions that occurred in the detector material before or inside the TPC tracking volume were reconstructed from the resulting charged particle daughters detected in the TPC. This was accomplished in three steps: the selection of track, pair, and primary photon candidates. All three steps utilized the unique topological signature of a photon conversion -- two tracks of opposite charge emerging from a secondary vertex with a small opening angle ($\approx$ $m_ec^2/E_\gamma$ radians, where $m_e$ is the electron mass and $c$ is the speed of light).

At the track level, improbable conversion daughters were removed by requiring tracks to satisfy a geometric cut and to have the ionization energy loss expected for an electron in the TPC gas. Neglecting resolution effects, the projection of a daughter track from a primary photon conversion onto the bend plane will form a circle which does not enclose the collision vertex. This is because photons typically propagate some distance before conversion and daughters emerge with a near zero opening angle (see Fig. \ref{fig::phot_con_schematic}). Thus, low {\pt} ($<$0.3{\gevc}) tracks with circular projections that enclosed the collision vertex in the bend plane of the 0.25 T solenoidal magnetic field were immediately removed. This cut was important, since the elimination of non electron (positron) tracks in this region of {\pt} via ionization energy loss {\dedx} is difficult due to the fact that the highly populated pion band crosses the electron {\dedx} band. It was not necessary to use this cut at higher {\pt}, since the yield of particles drops and electron (positron) identification via {\dedx} improves. It is also the case that at higher {\pt} this cut begins to remove daughters of primary photons since stiff track geometries make the distance from the collision vertex to the closest point on the circular projection of the helix comparable to the resolution of the measurement. At all momenta, electron and positron candidates which had a {\dedx} value between -2 and 4 standard deviations ($\sigma_{res}$) of the value expected for electrons and positrons were retained ($\sigma_{res}$ denotes the resolution of charge particle {\dedx} measurements in the TPC gas; $\sigma_{res} \approx$ 8.2\% of the {\dedx} value measured with a clean sample of electrons and positrons). The predicted energy loss curves for electrons, pions, kaons, and protons are shown as a function of rigidity in Fig. \ref{fig::dedxAll}. The {\dedx} requirement was chosen to be asymmetric, because on the lower side of the electron band other particle bands run in parallel and contamination is more prominent. On the upper side of the electron band other particle bands approach and cross the electron {\dedx} band in a narrow range of momentum. This reduced the usefulness of a tight cut on the upper side. It was estimated that approximately 3\% of the true photon daughters were removed with this {\dedx} requirement.

\beginFigure
\includegraphics*[width=.48\textwidth]{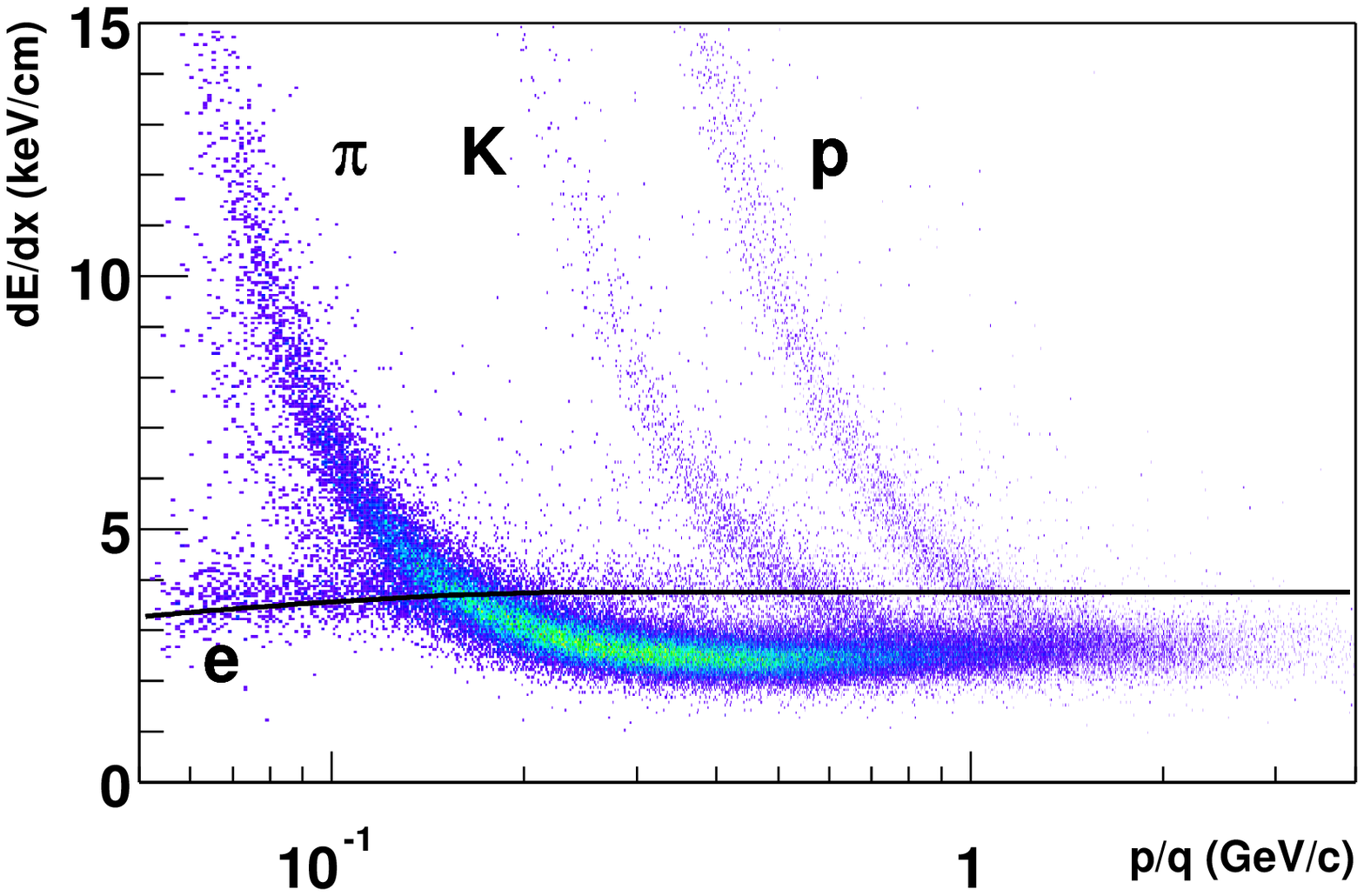}
\caption{Measured ionization energy loss in the TPC gas {\dedx} versus rigidity. Highlighted curves indicate the predicted {\dedx} curves for electrons, pions, kaons, and protons. Photon selection criteria have not been applied to track candidates in this figure.}
\label{fig::dedxAll}
\endFigure

Photon candidates were found by searching for track pairs which exhibited the topological signature of a photon conversion. Oppositely charged tracks were paired and passed through a geometric filter. The filter required each pair to originate from a secondary vertex with a near zero opening angle and low invariant mass. Secondary vertices were located by extrapolating daughter candidates to a common point. At the point of closest approach, daughters were required to come within 1{\cm} of each other in the non--bend plane ($rz$--plane) and within 1.5{\cm} of each other in the bend plane ($xy$--plane). The angular resolutions of opening angle measurements also differed in the two planes. In the non--bend plane and bend plane the precision of opening angle measurements have single Gaussian sigmas near 0.02 and 0.1 radians, respectively. Since even at energies as low as 100{\mev} photon conversions on average have an opening angle of 0.01 radians, ten times smaller than the precision in the bend plane, the full opening angle and the opening angle in the non--bend plane of each candidate pair were checked seperately. These values were required to be less than 0.4 and 0.03 radians respectively. The differing angular resolutions in the bend and non-bend planes are apparent in the invariant mass distribution of pairs assuming an electron (positron) hypothesis for the daughters (Fig. \ref{fig::phot_imass}). The invariant mass distribution has a sharp peak near zero and a broad peak close to 0.012\gevcc. The sharp peak at lower mass primarily results from cases where the bend plane projection of the opening angle was assumed to be zero. In these cases, the track geometries do not overlap in the bend plane, so the tracks were assumed to be parallel in the bend plane at the point of closest approach. Track geometries that do overlap in the bend plane lead to a higher invariant mass because the complete opening angle was used in calculation of the invariant mass. In this case, the less precise measure of the opening in the bend plane tends to dominate the result of the invariant mass calculation, moving and smearing the invariant mass peak. For this reason, a cut was placed on the value of the invariant mass of pairs calculated with only the non--bend plane projection of the opening angle. This cut required the invariant mass of candidate pairs to be less than 0.012\gevcc. The minimum mass returned by the calculation is $2m_e$ = 1.022\mevcc, which is above the first four 0.25{\mevcc} wide mass bins in Fig. \ref{fig::phot_imass}. This causes the absence of entries in the lower mass bins in the figure.

\beginFigure
\includegraphics*[width=.48\textwidth]{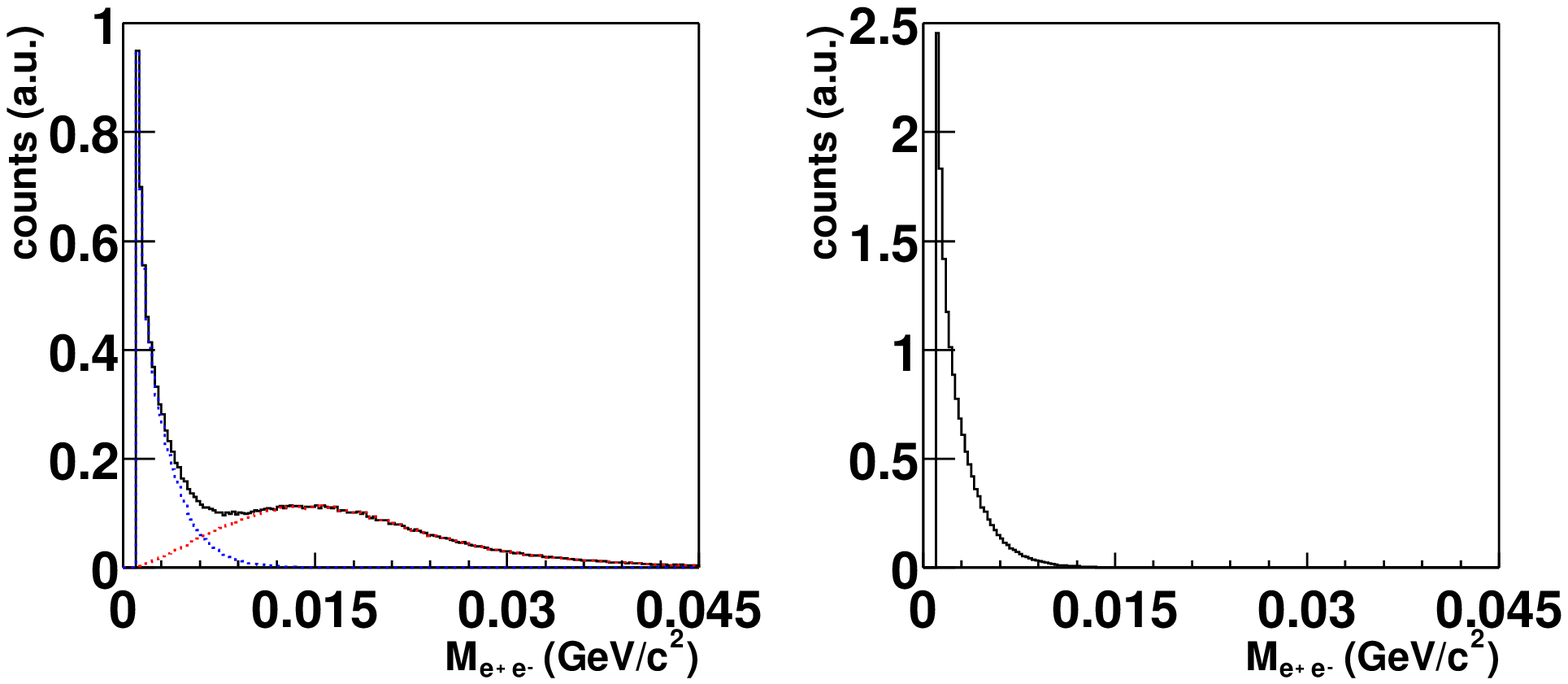}
\caption{Invariant mass distributions of photon candidates assuming the electron and positron mass for the daughters. Left figure: the distribution can be separated into two contributions (indicated with dashed lines); a sharp lower--mass peak primarily composed of track--pair geometries that do not overlap in the bend plane, and a higher invariant mass peak composed of pair geometries that do overlap in the bend plane. Right figure: the invariant mass distribution of photons calculated with the projection of the opening angle in the non--bend plane only.}
\label{fig::phot_imass}
\endFigure

The kinematic parameters for photon candidates were derived from the kinematic variables of the associated daughter tracks. The energy was calculated by summing the electron and positron energies. The angular direction in the bend plane was extracted by forming the cross-product of the vector from the helix center of the positron to the helix center of the electron, with the magnetic field vector. The angular direction in the non--bend plane was found by averaging the direction of electron and positron at the conversion point. With these three variables all kinematic parameters for photon candidates could be derived.

Primary photon candidates were selected from the set of all photon candidates by requiring the momentum vector to be consistent with the direction of a photon originating from the collision vertex. For these photons, the momentum vector has the same direction as the vector from the collision vertex to the conversion point. Due to differing angular resolution, the direction of the momentum vector and conversion point vector were compared separately in the bend and non--bend planes. For primary photon candidates, the difference between the momentum vector and the conversion point vector was required to be less than 0.035 and 0.015 radians in the bend and non--bend planes respectively. In order to reduce background in the photon sample from the random pairing of primary tracks, conversion vertices in the region close to the collision vertex ($r_{xy}$$<$10\cm) were excluded.

\subsection{Photon Spectra \label{sec::photSpec}}
Photon spectra were measured as a function of {\pt} and {\y} in three independent centrality classes as well as for an inclusive minimum bias data set. These spectra were produced from photon candidates identified with the standard event, track, and photon selection criteria (discussed in Sec. \ref{sec::reconPhot}).

Photon yields were extracted using the particle identification information of the positive daughters in the TPC. A {\dedx} deviant variable was constructed by comparing the energy loss predicted as a function of momentum for electrons and positrons with the measured {\dedx} and rigidity of daughter candidates, folding in the {\dedx} resolution of the TPC. This variable accounts for the momentum dependence in {\dedx}, and its value is therefore independent of the daughter particle's momentum and the parent photon's {\pt}. Consequently, the {\dedx} deviant values for daughters of differing momenta could be merged into bins based on the parent photon's {\pt} and {\y}. Distributions of the {\dedx} deviant values of positive daughters were chosen rather than the negative daughter to reduce the number of false photon candidates in the distributions that arise from knock-out electrons that originate when charged particles scatter in the detector material ($\delta$--electrons). The remaining contamination from this scattering process, which may result in a knock-out electron and positive particle ($\pi, K,$ or $p$) having a momentum in a region where the {\dedx} bands overlap the positron band (see Fig. \ref{fig::dedxAll}), was removed by requiring the fraction of the positive daughters energy to the total photon energy to be less than 75\%. The shape of the remaining background in the {\dedx} deviant distribution was studied on a sample of photon candidates that satisfied anti--photon cuts (primarily a sample of background candidates). Anti--photon cuts suppress positrons from true photons by requiring the photon selection criteria to be in the outermost extent of the cut distributions.  For example, the two--track distance of closest approach for daughter tracks was required to be 1.5$<$$|$d$_{xy}$$|$$<$2{\cm} and 1$<$$|$d$_{z}$$|$$<$1.5{\cm}. A two parameter exponential plus linear function was used to describe these background distributions,  as shown for one {\pt} bin in Fig. \ref{fig::photYieldHow}. Parameters of the background functions were found by fitting to the background distributions with the region around the expected value removed ({\dedx} deviants between -1.5 and 3 $\sigma_{res}$). This was necessary to avoid fitting the signal from residual photons that still existed after the application of anti--photon cuts. With the knowledge of the background shape, the raw yield of photon candidates was extracted using a three parameter Gaussian function plus the background function which had one free scaling parameter (also shown in Fig. \ref{fig::photYieldHow}).

The purity of the photon candidate sample was determined by dividing the integral of the Gaussian function by the integral of the entire Gaussian plus background function between {\dedx} deviant values of -2 and 4 $\sigma_{res}$. For {\pt} $<$ 0.75{\gevc}, the purity of the photon candidate sample is greater than 90\% in all centrality classes. In the 0--11\% most central centrality class, where the purity is the lowest, the purity drops linearly from approximately 90\% at {\pt} = 0.75{\gevc} to about 60\% at {\pt} = 2.4\gevc. A cleaner sample (purity $>$95\% below {\pt} = 0.90{\gevc} for the 0--11\% most central collisions) was obtained by requiring photons to convert in the inner field cage and TPC gas, $r_{xy}$$>$40\cm.

Uncorrected yields were obtained from the weighted sum of the entries in the distributions. The weights were extracted by dividing the height of the Gaussian function by the height of the entire fit at the location of each entry. Distributions in {\y} were extracted in a 4 x 10 array of {\pt}--{\y} bins to properly account for the variation in efficiency as a function of {\pt}. Three 0.25{\gevc} wide {\pt} bins were used below {\pt} = 0.75{\gevc} where the efficiency grows rapidly, and one large bin was used for 0.75$<${\pt}$<$2.5{\gevc} where the efficiency is flat. The {\pt} distributions did not require division into {\pt} and {\y} bins, since the corrected {\y} distributions and the input distributions for the efficiency calculations were both uniform in {\y} for $|$\y$|$$<$0.5.

\beginFigure
\includegraphics*[width=.48\textwidth]{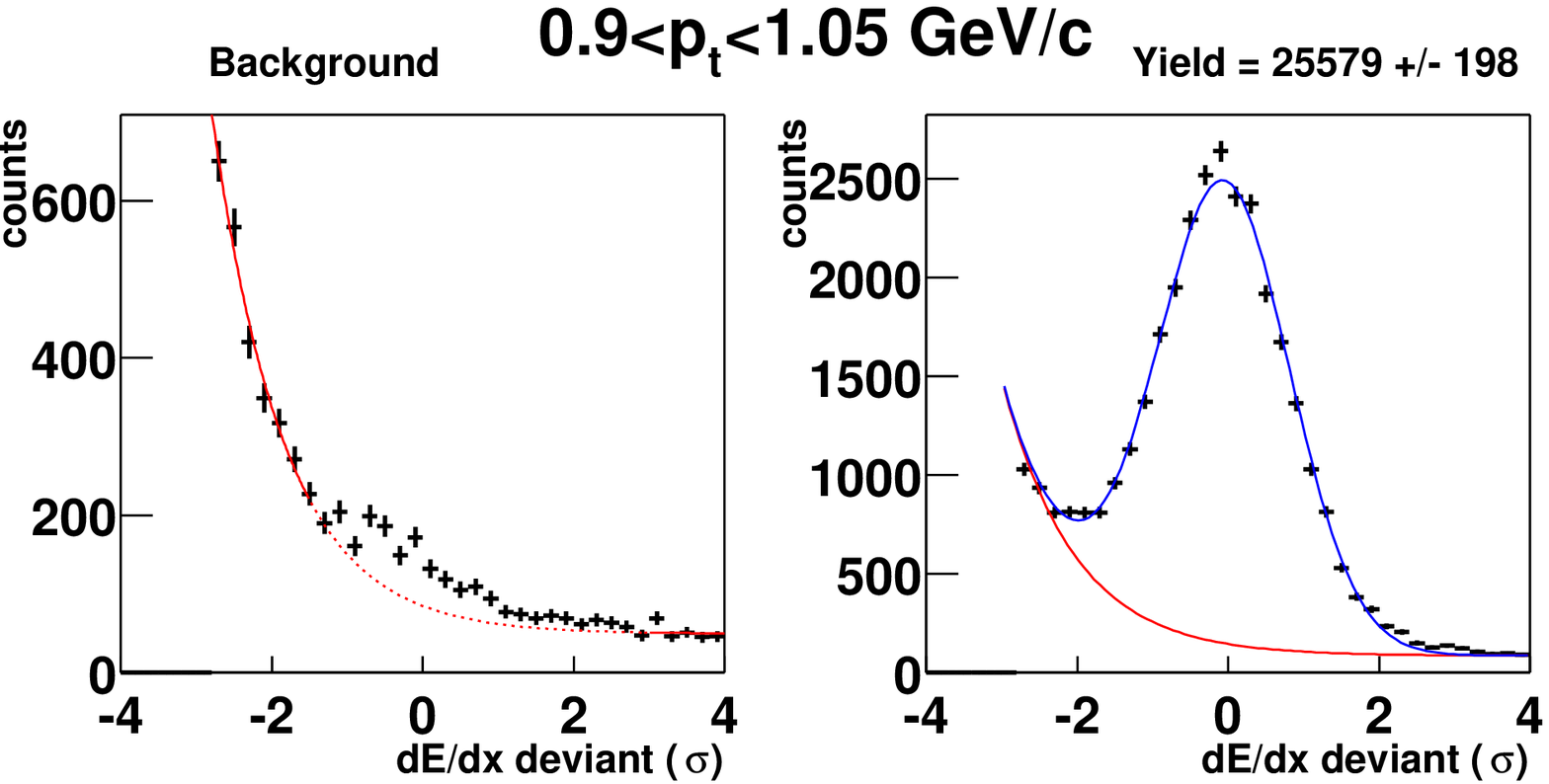}
\caption{{\dedx} deviant distributions of positive daughters with 0.9$<${\pt}$<$1.05{\gevc} from photon candidates with $r_{xy}$$>$10{\cm}. Left figure: distribution of background candidates fit with an exponential plus linear function. Right figure: distribution of the positron signal fit with a Gaussian function plus a scaled background function.}
\label{fig::photYieldHow}
\endFigure

Efficiency corrections were applied to each {\pt}--{\y} bin independently. These corrections were calculated with detailed simulations (GEANT 3.21) of the propagation of photons and daughter particles through a realistic detector geometry. A TPC Response Simulator (TRS) was used to simulate the drift and electronic response of ionization deposited in the TPC. Digital pad signals produced by TRS were embedded pixel--by--pixel into real events.

Each simulated event contained 2000 photons generated flat in {\pt} and {\y}. On average, approximately 20 of these photons interacted with the detector material (a consequence of the low conversion probability). This added the ionization of about 40 daughter tracks to each real event. This number is less than 2\% of the number of charged particles in the embedded phase space in a typical high multiplicity event. Therefore even in high multiplicity events, which are most sensitive to over--embedding, the introduction of 2000 photons into each event had a negligible effect on the track reconstruction efficiency. An association process was used to link reconstructed and generated photons. The photon finding efficiencies for different centrality definitions were calculated by dividing the distributions of reconstructed photons correctly associated with a generated photon by the input distributions of generated photons.

\beginFigure
\includegraphics*[width=.48\textwidth]{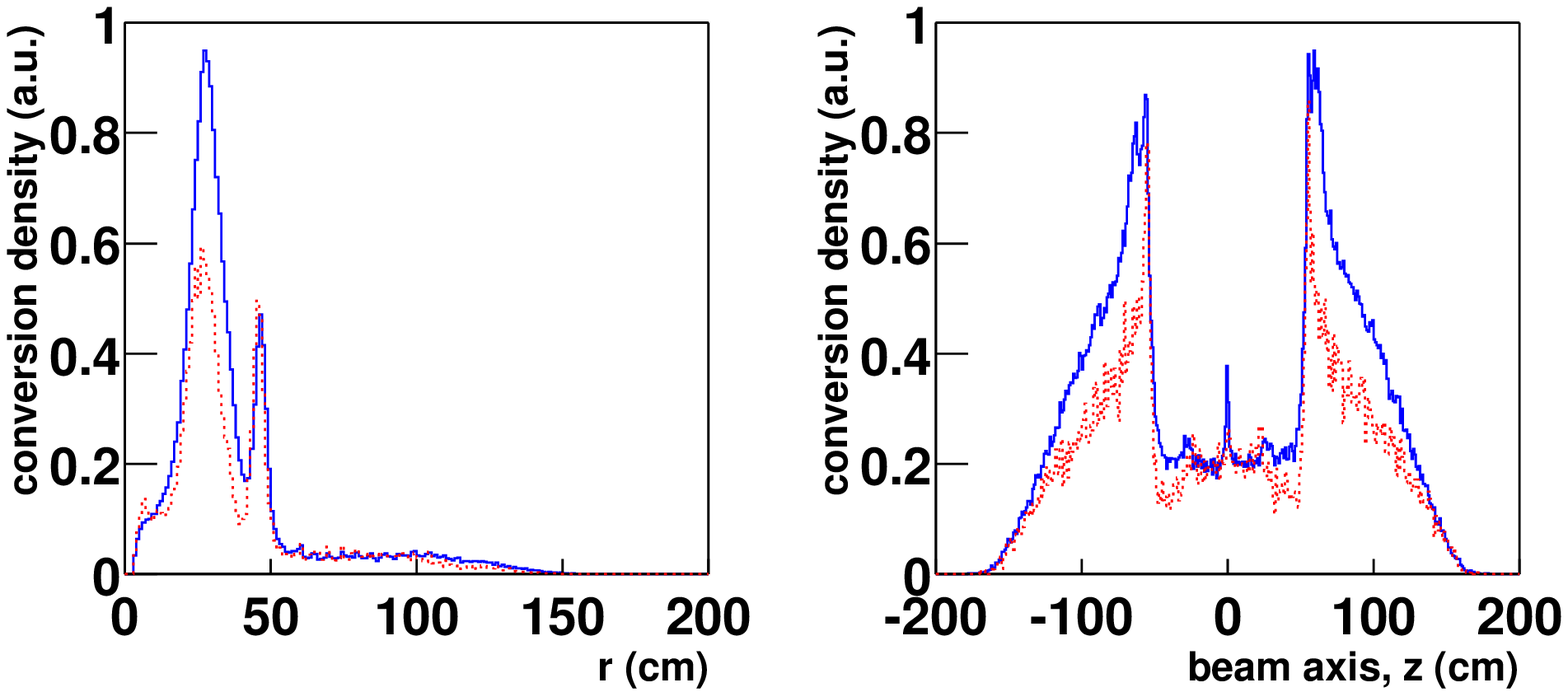}
\caption{Number of reconstructed photon conversions as a function of the conversion location for both real data (solid line) and GEANT simulated events (dashed line). Left figure: conversion density as a function of radial distance from the beam axis. Right figure: conversion density as a function of the distance along the beam axis, $z$.}
\label{fig::materialDiff}
\endFigure

To reveal systematic trends caused by differences in the layout of the detector material and the material map used in simulation to calculate efficiency corrections, spectra were produced with different requirements on the minimum distance between the location of conversion and the beam axis in the $xy$--plane ($r_{xy}$$>$10{\cm} and $r_{xy}$$>$40{\cm}). It was found that an additional correction factor was needed to compensate for differences in the layouts of the detector material between $r_{xy}$ = 10{\cm} and $r_{xy}$ = 40{\cm}, as illustrated in Fig. \ref{fig::materialDiff}. Above $r_{xy}$ = 40{\cm}, the material of the inner field cage and the gas of the TPC were well described in the simulation. This region was used as a reference to calculate the correction factor (1.42) for the $r_{xy}$$>$10{\cm} spectra. All {\pt} and {\y} data points in the $r_{xy}$$>$10{\cm} spectra were linearly scaled by this factor.

Corrected {\pt} and {\y} spectra for the various centrality classes are shown in Fig. \ref{fig::photSpectra}. Systematic uncertainties of 7\% point--to--point and 12\% overall (correlated) have been estimated for the measurements in the {\pt} spectra. There is a 12\% (correlated) systematic uncertainty in the normalization of the {\y} spectra. These uncertainties account for uncertainty in the detection efficiency and potential measurement bias that may arise from differences in the physical and simulated material maps.

\beginFigure
\includegraphics*[width=.48\textwidth]{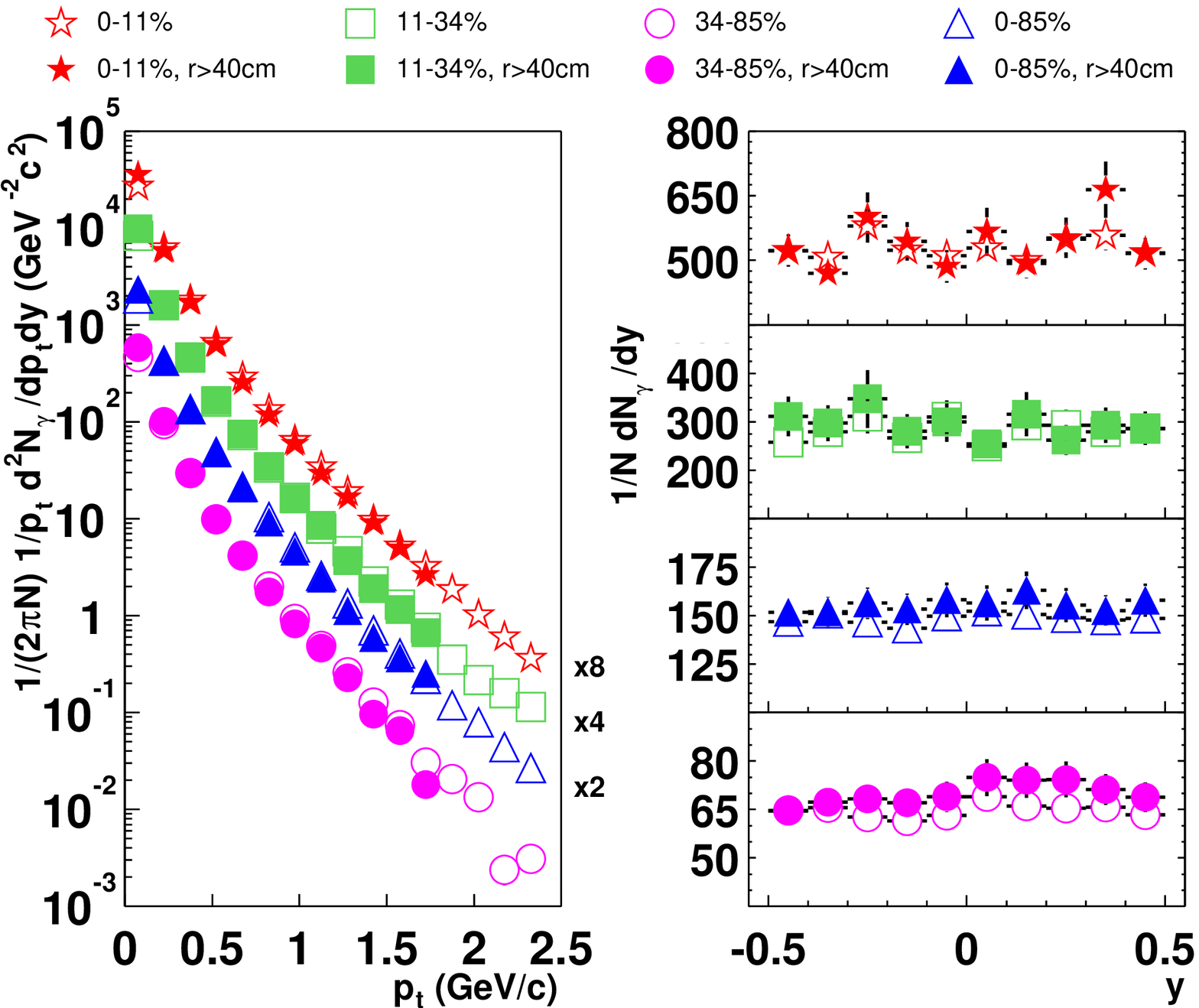}
\caption{Corrected photon {\pt} (left) and {\y} (right) spectra for {\auau} collisions at {\sqnn} = 130\gev. The {\pt} distributions are for mid--rapidity photons, $|$\y$|$$<$0.5. Statistical uncertainties are shown. Systematic uncertainties in the {\pt} spectra and on the normalization of the dN/d{\y} spectra have not been included in this figure.}
\label{fig::photSpectra}
\endFigure

Values in the lowest $r_{xy}$$>$10{\cm} {\pt} bins, 0$<$\pt$<$0.15{\gevc}, are systematically 15--25\% lower than in the corresponding $r_{xy}$$>$40{\cm} bins. This is attributed to the efficiency correction being underestimated in these bins for the $r_{xy}$$>$10{\cm}, because the simulation lacked material between 10$<$$r_{xy}$$<$40\cm. This gave the inner field cage and TPC gas a larger fraction of the total conversion probability in the simulation, and resulted in a mean $r_{xy}$ conversion point that is closer to the TPC. The combination of the shift in the mean position of conversions and the linear scaling of the spectra to compensate for differences in the material layouts artificially increased the efficiency of low {\pt} photons in the $r_{xy}$$>$10{\cm} spectra. Therefore the corrections, which are the inverse of the efficiencies, are too small in the $r_{xy}$$>$10{\cm} spectra at low {\pt}. 

\subsection{Measuring the yield of {\pizgg} decays\label{sec::pizFitting}}
The uncorrected yield of {\piz} mesons was extracted from the invariant mass distributions of photon pairs in various {\pt} bins. Individual decays could not be uniquely identified, because of the large combinatorial background in the invariant mass distributions. The combinatorial backgrounds were simulated by  {\it rotating} the momentum vector of one photon in each pair by $\pi$ radians in the bend plane. In this way, it was possible to create combinatorial background distributions which preserved event characteristics such as the vertex position along the beam axis, the event multiplicity and anisotropic flow. Due to the azimuthal symmetry of the STAR TPC, this type of rotation also ensured that a consistent geometric acceptance and track reconstruction efficiency were maintained. At the same time these rotations moved and smeared the invariant mass values of the pairs that are correlated through two photon decays. The shape of the resulting background distributions near the {\piz} mass ($\pm$0.1{\gevcc}) was well described and smoothed with a second order polynomial (see Fig. \ref{fig::pi_imass}). This functional form was also used to describe the shape of the combinatorial background in the {\it unrotated} invariant mass distributions. A Gaussian function was used to describe the enhancement at the {\piz} mass. The width ($\sigma$) of the enhancement ranged between 4 and 15{\mevcc}, and was found to be consistent with simulations in all centrality classes and as a function of {\pt}. It was dominated by the photon momentum resolution not the intrinsic mass width of the {\piz} ($\sim$8\evcc). For this reason a Breit--Wigner function that would be appropriate to describe the intrinsic width of such a resonance was not used. Equation (\ref{eq::pi0_equation}) is the complete function that was used to describe invariant mass distributions of photon pairs. Systematic uncertainty related to the choice of the invariant mass bin width, $\delta$, was studied by comparing the measured yields for various choices of $\delta$. These studies indicated that the systematic uncertainty due to this effect was much smaller than statistical uncertainties in the measurements. 

\begin{equation}
  C\left(x{\equiv}M_{\gamma\gamma}\right) = \frac{N\delta}{\sigma\sqrt{2\pi}}e^{-\left(x-m\right)^2/\left(2\sigma^2\right)} + B\left(a+bx+cx^2\right),
\label{eq::pi0_equation}
\end{equation}
{\EqTextStyle{where $\delta$ is the width of the invariant mass bins, N is the number in the Gaussian peak, and B is the scale factor of the background function.}\\}

\beginFigure
\includegraphics*[width=.25\textwidth]{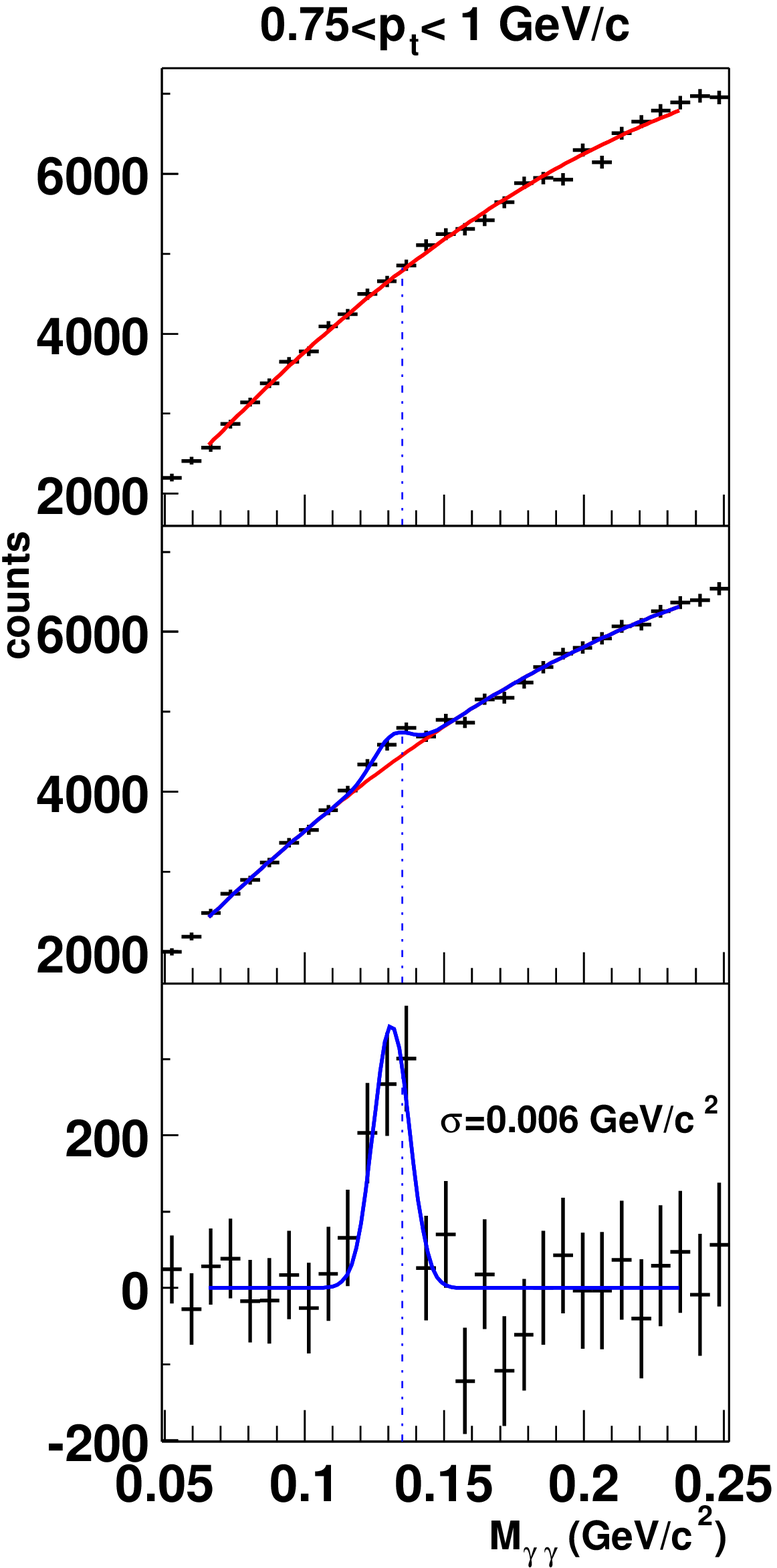}
\caption{Two--photon invariant mass distributions for candidates with 0.75$<$\pt$<$1{\gevc} in the 0--11\% most central {\auau} collisions at {\sqnn} = 130\gev. Top frame: invariant mass distribution with one photon {\it rotated} by $\pi$ radians fit with a second order polynomial. Middle frame: invariant mass distribution of photon pairs fit with Eq. (\ref{eq::pi0_equation}); the background function is also shown. Bottom frame: invariant mass distribution after the combinatorial background was removed. The enhancement near the {\piz} mass is located at 0.131{\gevcc} and has a Gaussian sigma of 0.006{\gevcc}}
\label{fig::pi_imass}
\endFigure

A feature of this method of photon reconstruction is that the location of the enhancement from {\piz} decays in the two-photon invariant mass distribution is a few MeV/c$^2$ lower than expected (see Fig. \ref{fig::pi_imass}). This is attributed to energy loss experienced by the electrons and positrons in the detector material. The ``global'' tracking routine used in this analysis only compensated for energy loss in the TPC gas and not for that in other detector material. This resulted in the reconstructed momentum for electrons and positrons originating prior to the gas volume of the TPC to be systematically lower than their original momentum. The small ({$\sim$1\mev} on average) energy loss experienced by each of the 4 daughter particles translated to a few MeV/c$^2$ shift in the location of the {\piz} invariant mass peak. This hypothesis is consistent with a similar feature in simulated events. The location of the reconstructed invariant mass peak for simulated {\piz}s systematically decreased as the radial distance between the beam axis and conversion point of the closer photon in the pair decreased. This implies that a larger {\piz} mass deviation occurs when more detector material is between the conversion point and the TPC. The reconstructed energy of simulated photons was also systematically lower than the energy input to the simulation and larger deviations also occurred when more detector material was between the conversion point and the TPC.

\subsection{Spectra of {\piz} mesons\label{SecPiSpectra}}
Two iterations (described in Sec. \ref{sec::pizFitting}) were performed to fit the data and extract the yield of {\piz}s as a function of {\pt}. For the first iteration, 4 free parameters were used in the fit: the yield (N), mass (m), and width ($\sigma$) of the Gaussian function describing the peak, and the scale factor for the background function (B). For the second iteration it was assumed that the width of the invariant mass peak increased linearly with {\pt}, as seen in the simulated events. The width parameters for the fits in the second iteration were obtained from a linear fit to the widths found in the first iteration. The values of the slopes for these linear fits ($\approx$ 3$\left(MeV/c^2\right)/\left(GeV/c\right)$) were consistent with those found in simulation. The width parameters were then fixed to the value of the linear function at the center of each {\pt} bin. This reduced the number of free parameters in the second pass to three and increased the stability of the fits.

Uncorrected yields about mid--rapidity ($|$\y$|$$<$1) were extracted in various {\pt} bins for the four different centrality classes. The narrow 6{\mevcc} width (sigma) of the enhancement at the {\piz} mass measured with this reconstruction method is significantly better than what is typically achieved using a conventional lead-scintillator sampling calorimeter (20{\mevcc} sigma). The improvement is a result of the excellent photon energy resolution (3\% at 1\gev) obtained with this method of photon reconstruction. The narrow width improves the signal to background ratio and enables the extraction of raw {\piz} yields at low {\pt} (\pt$<$0.75\gevc) where the signal to background ratio is seriously degraded by a large combinatorial background. 

Efficiency corrections were calculated with a procedure similar to the one used to calculate the photon detection efficiencies (described in Sec. \ref{sec::photSpec}), except that {\piz}s were selected in GEANT. Only the ionization in the TPC gas from daughters of those {\piz}s selected in GEANT was passed to TRS. This was necessary to perform the calculation in a reasonable amount of cpu time and to prevent saturating real events with ionization. One consequence of the low conversion probability is that on average only 1 in 10000 {\pizgg} decays is expected to be detected through the reconstruction of pair conversions. At the same time about 1 in 50 {\pizgg} decays produces a photon that converts to create a pair of tracks that ionize the gas in the TPC. Selecting detectable {\pizgg} decays in GEANT ({\piz}s that decay into two photons, with both photons undergoing interactions which create at least one daughter within the TPC acceptance) reduced the amount of uninteresting ionization in each simulated event. With this selection, the insertion of up to 12 detectable {\pizgg} decays into each event added less than 2\% to the number of tracks in the phase space of the embedding. These events were reconstructed with the same software used to reconstruct real events. Reconstructed photons that could be associated with a photon from a simulated {\pizgg} decay were retained. These photons were used to generate two-photon invariant mass distributions. The yields of simulated {\piz}s were calculated in the same {\pt} and centrality bins, and with the same fitting procedure as the raw yields. Efficiency corrections for these bins were obtained by dividing the reconstructed distributions by the input distributions.

Corrected {\pt} spectra of {\piz}s were obtained by applying efficiency corrections to the {\pt} distributions of the raw yields. The corrected spectra are shown in Fig. \ref{fig::pi0corrYields}. The uncertainties shown are statistical and mainly reflect the low number of real {\piz}s measured. They combine the uncertainty in the raw yields and efficiency corrections. Systematic uncertainties due to the cuts used were studied by varying track, photon, and {\piz} cuts. These studies revealed that the statistical fluctuations dominate the systematic variations in this analysis.

\beginFigure
\includegraphics*[width=.44\textwidth]{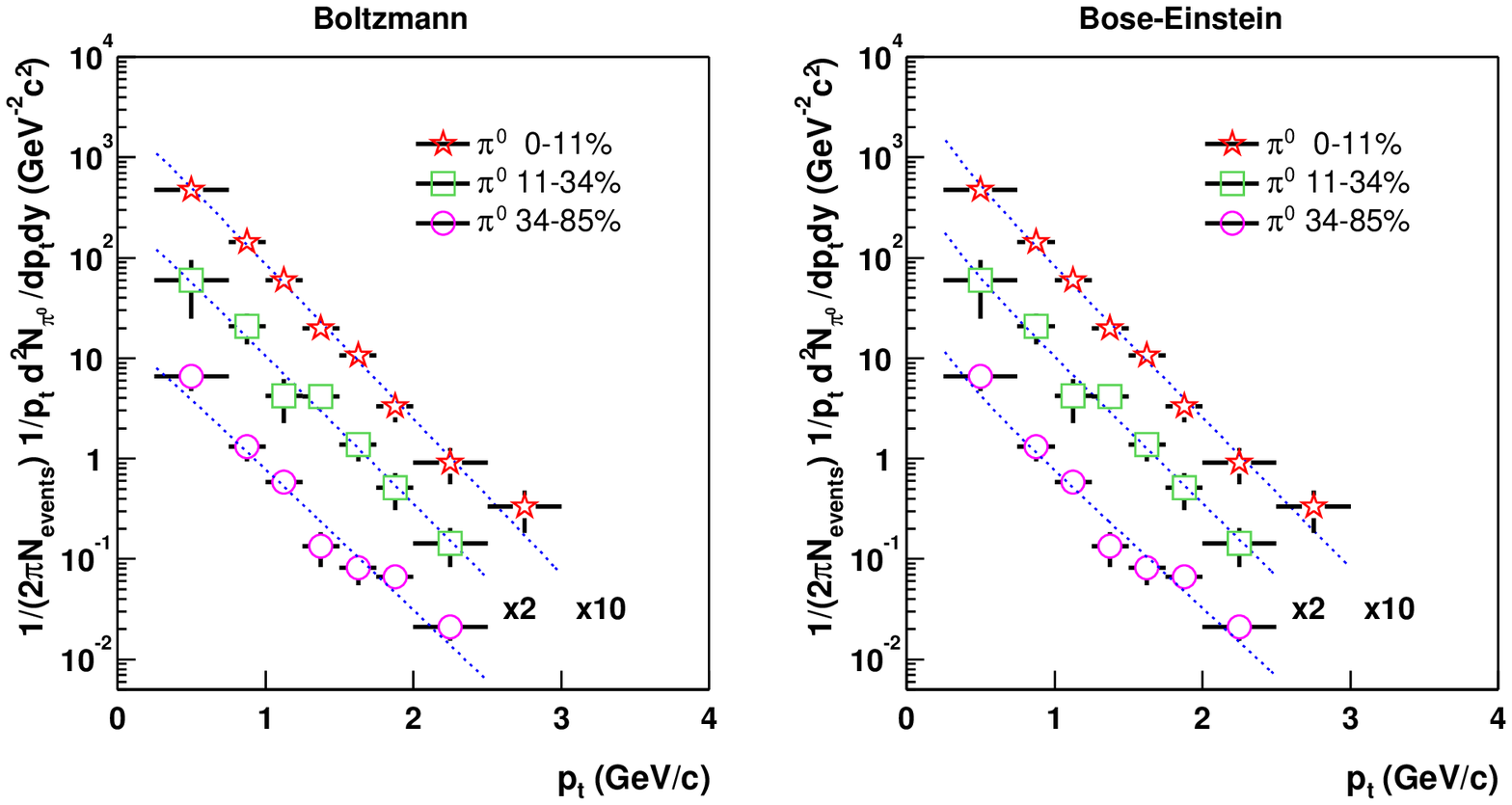}
\caption{Transverse momentum spectra of the {\piz} meson about mid--rapidity, $|$\y$|$$<$1, for different centrality bins. Along with the point--to--point statistical uncertainties shown, the spectra have a common uncertainty in the normalization of {$\pm$}40\%. The central data (0--11\%) has an additional normalization uncertainty of {$\pm$}19\%, which results in a total normalization uncertainty of {$\pm$}49\%. Dashed lines indicate Boltzmann fits to the spectra.}
\label{fig::pi0corrYields}
\endFigure

Corrections to the normalization of the spectra were made to compensate for the overall difference in the photon conversion probability in the real detector material and that used in simulation. These correction factors were based on the corrected yield of {\piz}s (Y$_{r_{xy}>40}$) for photons that converted in the inner field cage or the TPC gas ($r_{xy}$$>$40\cm) where the two material maps are consistent with each other (see Fig. \ref{fig::materialDiff}). For 1$<${\pt}$<$2\gevc, a centrality--independent material correction factor was directly extracted by dividing Y$^{min}_{r_{xy}>40}$ by Y$^{min}$ for minimum bias events. In this {\pt} interval, the ratio of the {\piz} efficiency for Y$^{min}_{r_{xy}>40}$ to the efficiency for Y$^{min}$ is uniform. Therefore it was not necessary to correct this factor for interplay between the shape of the efficiency correction and the exponentially falling spectrum. This material correction factor was crosschecked in three different {\pt} windows. The resulting variation in the the factors was found to be 5$\times$ smaller than the uncertainty in an individual factor ($\sim$40\%). This uncertainty is statistical and arises due to the relatively small number ($\sim$50) of {\piz}s reconstructed from photons with $r_{xy}$$>$40{\cm}. The stability of this measurement was checked by comparing the results for different invariant mass bin widths. This check confirmed that the statistical uncertainty in the raw yields is the dominant uncertainty. The material correction factor for the minimum bias data set was 0.31 {$\pm$} 0.12, with 1$<${\pt}$<$2{\gevc} as the window of {\piz} \pt. This factor was used to scale the normalization of centrality classes formed by taking subsets of the minimum bias triggered data set (0--85\%, 34\%--85\%, 34\%--11\%). The uncertainty in this factor ({$\pm$}0.12 or 40\%) is common for all these centrality bins and cancels out when ratios are taken. This factor can not be used to normalize the 0--11\% centrality bin, because of the differing $z$ vertex distributions between the central and minimum bias triggered data sets. A separate centrality--independent factor was calculated for $|z_{vertex}|$$<$75{\cm} where the vertex distributions are similar in the two data sets. Minimum bias triggered events with $z$ vertices in this region were used to calculate the factor. The factor, 0.26 {$\pm$} 0.11, was obtained by dividing Y$^{min}_{r_{xy}>40}$ by Y$^{min}_{|z_{ver}|<75}$. Using this factor, the normalized yield for the 0--11\% centrality bin was computed by taking the product of the factor and Y$^{cent}_{|z_{ver}|<75}$. This normalized yield, Y$^{cent}$ = 12.0 {$\pm$} 5.7 for 1$<${\pt}$<$2\gevc, is independent of the vertex distribution. The final 0--11\% spectrum was scaled by Y$^{cent}$/Y$^{cent}_{|z_{ver}|<150}$ = 0.27 {$\pm$} 0.14. In summary, the 0--11\% most central class of events has the same common uncertainty of {$\pm$}40\% plus an additional uncertainty of {$\pm$}19\%. These two uncertainties were combined ({$\pm$}49\%) for the purpose of comparison to other spectra, like the spectrum of charged hadrons.

\section{Contribution of the {\pizgg} decay to the inclusive photon spectrum}
Electromagnetic decays of neutral mesons are the dominant source photons in heavy ion collisions. Among these, the {\pizgg} decay is largest contributor to the spectrum of inclusive photons. Its large contribution hides the signal from other sources, such as direct photons emitted during the early stages of heavy ion collisions.

To investigate how the {\pizgg} decay contributes to the inclusive photon spectrum, the {\pt} distributions of {\piz}s were used to generate the single photon spectrum expected for the daughters of {\pizgg} decays. The {\pt} distributions of {\piz} were fit with both a Boltzmann function and a Bose--Einstein function. For both functions, the total energy of the {\piz} was replaced by its transverse energy ($\sqrt{p_t^2c^2+m_\pi^2c^4}$) under the assumption that the system is boost invariant near mid--rapidity. This assumption is supported by the flat shape of particle rapidity distributions close to mid--rapidity \cite{star_hadron,phobos_rapidity}. Other more sophisticated functions, incorporating resonances that decay into {\piz}s and/or handling radial expansion of the system in more detail with additional parameters, were not chosen because the additional parameters were not well constrained by the 7 or 8 data points of the spectra. Both the Boltzmann and Bose--Einstein functions treat the system as a thermalized gas and converge to exponential functions at high \pt.

Distributions of the {\pt} dependence of {\piz}s were generated using these functions, assuming that the rapidity and azimuthal distributions are flat. The input rapidity distribution of {\piz}s was limited to $|$\y$|$$<$2. This rapidity window produces more than 99\% of the photons with $|$\y$|$$<$0.5 from {\pizgg} decays. These distributions were passed through a Monte Carlo decay simulator used to calculate the {\pizgg} decay kinematics and boost between the center of momentum and laboratory frames. The momentum information of the decay photons was used to produce the single photon {\pt} spectra of the daughters.

The fraction of the photons from {\pizgg} decays in the inclusive photon spectra was calculated as a function of {\pt} by dividing the simulated spectra by the measured inclusive photon spectra, as shown in Figure \ref{fig::photonDiff}. The shape of the resulting distributions for various centrality classes is independent of the uncertainty in the normalizations of the {\piz} spectra, but may depend on the assumed {\pt} dependence of the {\piz} spectra. For this reason, the fractions are shown for \pt$>$0.45{\gevc}, where the photon contribution is determined from {\piz}s in and above the measured {\pt} interval. The kinematics of {\pizgg} decay limit the {\pt} of the daughter photons. For example a {\piz} of \pt$<$0.435{\gevc} can only produce photons with {\pt} $<$ 0.45\gevc. Thus, the unmeasured portion of the {\piz} spectra below {\pt} = 0.25{\gevc} does not contribute to the photon spectra in the region where the fractions are plotted.

The fraction of photons from {\pizgg} decay in the inclusive photon spectrum is approximately constant between 0.75$<$\pt$<$1.65{\gevc}. For the 0--11\% most central event class, the fraction begins to decrease substantially near {\pt} = 1.65{\gevc} assuming either the Boltzmann function or Bose--Einstein function to describe the {\pt} spectra of {\piz}s. Specifically, the relative contribution from {\pizgg} decays decreases by 20\% $\pm$ 5\% from {\pt} = 1.65{\gevc} to {\pt} = 2.4\gevc. Both point--to--point uncertainties in the photon spectrum as well as substantial uncertainty in the slope parameters of the Boltzmann (0.281{\gev} $\pm$ 0.013) and Bose--Einstein (0.289{\gev} $\pm$ 0.014) fits have been included in the 5\% uncertainty in the region of the decrease. A similar trend was observed by the WA98 collaboration in {\pbpb} collisions at {\sqnn} = 17.2{\gev} \cite{WA98_direct_photons,WA98_long}. The WA98 collaboration reported an excess of photons above 1.5{\gevc} in central collisions after accounting for photons from all expected electromagnetic decays. For electromagnetic decays other than the {\pizgg} decay, the $\eta\rightarrow\gamma\gamma$ decay channel is expected to be the next largest contributor. From {\pt} = 1{\gevc} to {\pt} = 4{\gevc}, its contribution is expected to be approximately 15\% and to be fairly uniform in {\pt}, increasing less than 5\% per\gevc. The WA98 collaboration has estimated that the summed contribution of all other electromagnetic decays is less than a few percent at SPS energies \cite{WA98_long}. Based on the above assumptions (the {\piz} {\pt} spectrum has a Boltzmann or Bose--Einstein {\pt} dependence, the $\eta$ contribution increases by less than 5\% per{\gevc}, and the summed contribution of all other electromagnetic decays is less than a few percent) it is unlikely that electromagnetic decays fully account for the observed single photon yields in the 0--11\% most central event class.

\beginFigure
\includegraphics[width=.44\textwidth]{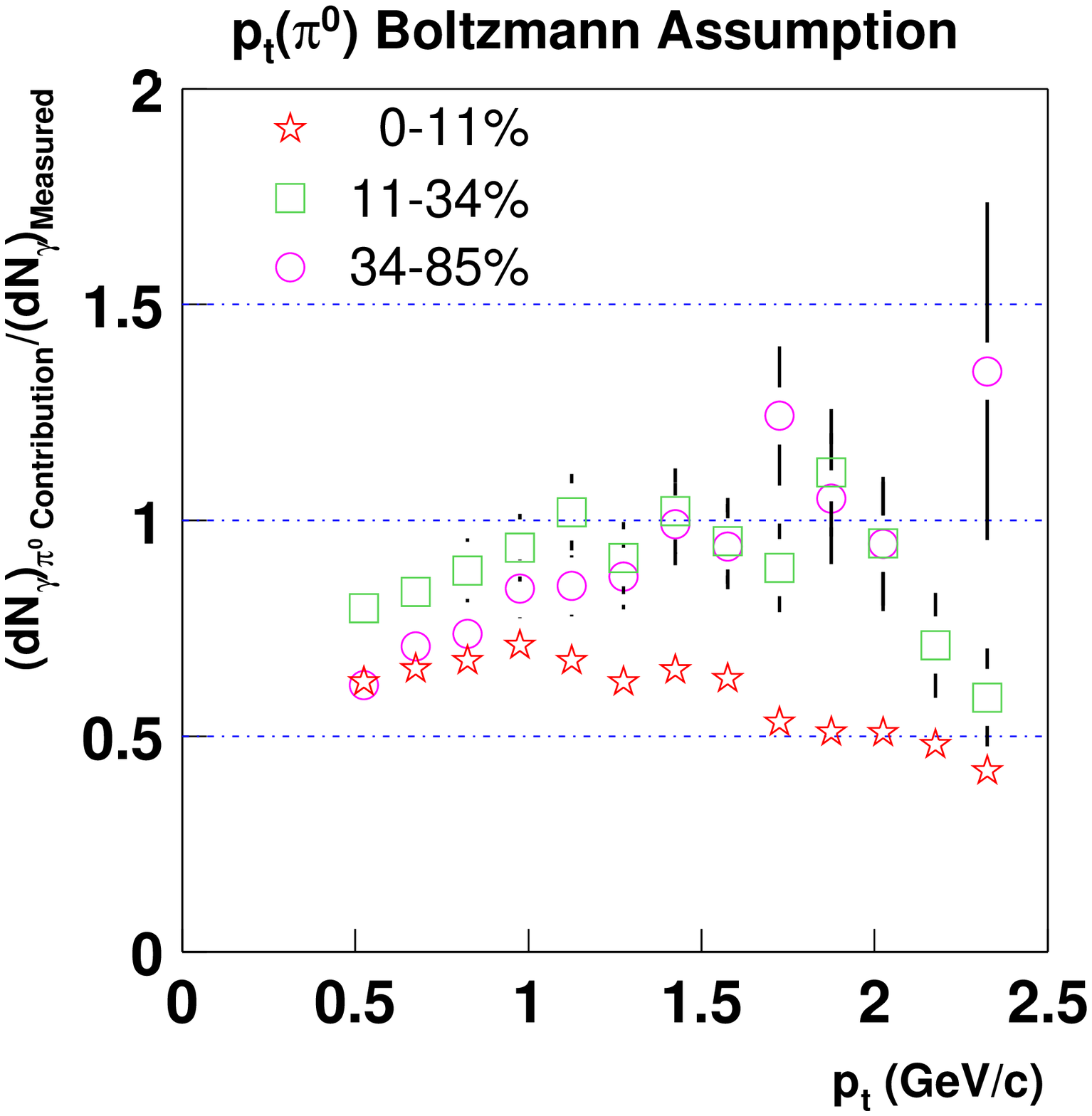}
\caption{The Ratio as a function of {\pt} of the distributions of photons from {\pizgg} decays to the measured photon spectra. The {\pizgg} photon distributions were generated assuming that the {\pt} dependence of the {\piz} {\pt} spectra follow a Boltzmann distribution. These ratios include both statistical and systematic uncertainties in the photon spectra. Uncertainties in the normalization of the {\piz} spectra arise as {\pt} independent uncertainties and have not been included. Normalization uncertainties in the {\piz} spectra are 40\%, correlated between all ratios, with an additional 19\% uncorrelated uncertainty for the 0--11\% centrality ratio. Uncorrelated {\pt} dependent uncertainties that arise from the uncertainty in the slope parameters of fits to the {\piz} {\pt} spectra (11\%, 9\% and 5\% respectively for the 34--85\%, 11--34\% and 0--11\% centrality classes) have not been included.}
\label{fig::photonDiff}
\endFigure

\section{Comparisons to Published Data}
Comparisons between the 0--11\% {\piz} spectrum and the 0--10\% charged hadron spectrum were used to study the composition of the hadron spectrum as a function of {\pt}. The ratio of the {\piz} data points to a power law function fit to the charged hadron spectrum ($\left(h^-+h^+\right)/2$) is shown in Fig. \ref{fig::hadComp}. At {\pt} = 2{\gevc} the ratio of {\piz}s to charged hadrons approaches 50\% (also shown in Fig. \ref{fig::hadComp}). Assuming isospin symmetry for charged and neutral pions ($\left(dN_{\pi^+} + dN_{\pi^-}\right) \equiv 2\left(dN_{\pi^0}\right)$) the proton to pion ratio is close to 1 at {\pt} = 2{\gevc}. This result is similar to the previous observation in central collisions that the ``$\bar{p}$ and $p$ yields are comparable to the $\pi^{+/-}$ yields.'' \cite{phenix_charged}. Another method of probing the ratio of baryons to mesons in the system is by examining the ratio of {\piz} to $\Lambda$ production (shown in Fig. \ref{fig::lambdaComp}). In this figure, Bose--Einstein functions have been used to describe the $\Lambda$ and $\bar\Lambda$ spectra. The value of the {\piz} data points has been divided by the value of the Bose--Einstein function describing the $\Lambda$ and $\bar\Lambda$ spectra at the center of the bins to obtain the {\piz} to $\Lambda$, and {\piz} to $\bar\Lambda$ ratios. At {\pt} = 2 {\gevc} these ratios are approximately 1, consistent with other measurements of the baryon to meson ratio.

\beginFigure
\includegraphics*[width=.48\textwidth]{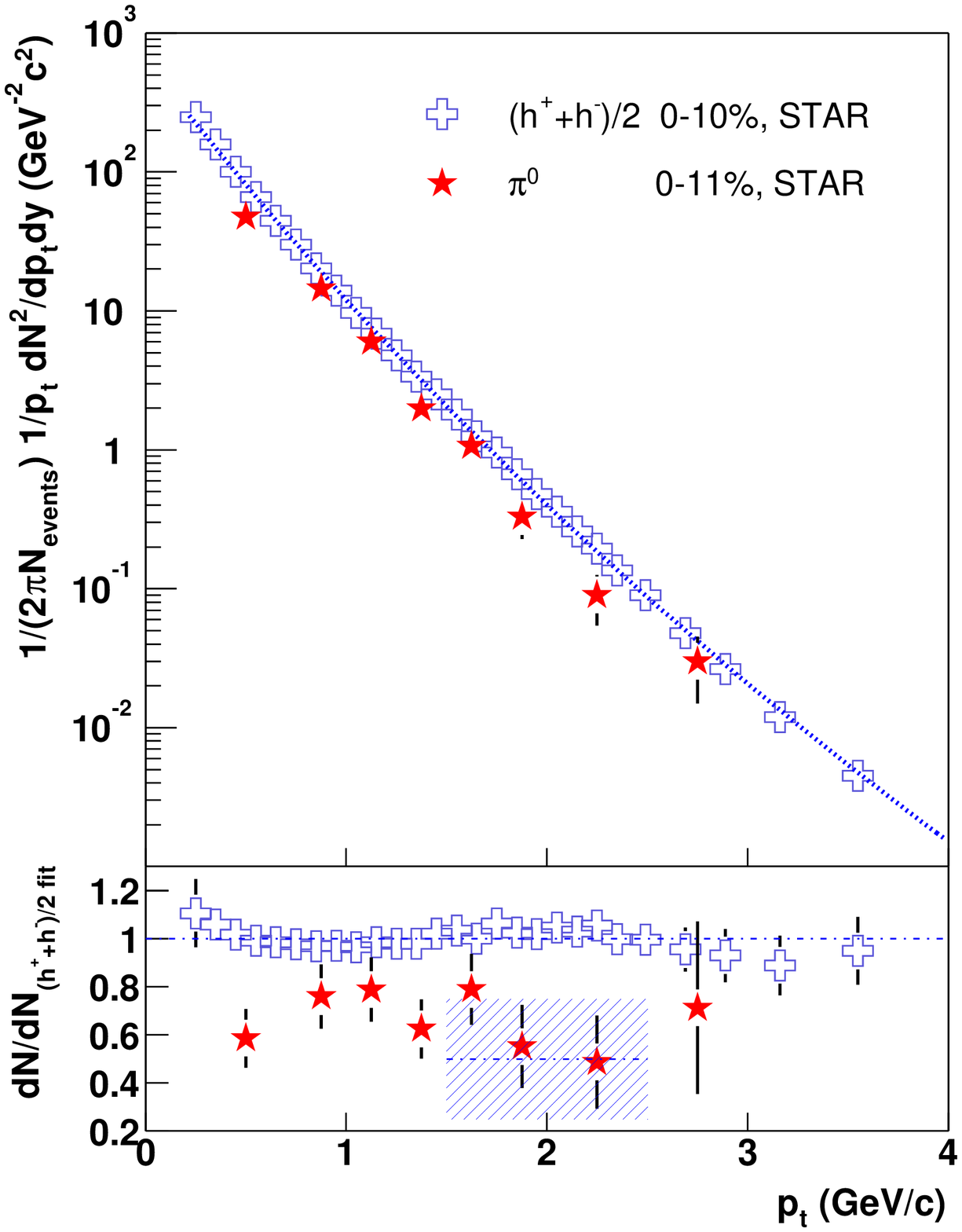}
\caption{Top frame: comparison between STAR {\piz} and inclusive charged hadron ($\left(h^-+h^+\right)/2$) spectrum \cite{star_highpt} about mid--rapidity for {\auau} collisions at {\sqnn} = 130\gev. Bottom frame: ratios of these spectra to a power law fit to the STAR inclusive charged hadron spectra. For reference, dashed lines indicate $dN/dN_{\left(h^-+h^+\right)/2~fit}$ of 0.5 $\pm$ 0.25 and 1. Normalization uncertainties in both the {\piz} ($\pm$49\%) and inclusive charged hadron ($\pm$11\%) measurements have not been included with the data points.}
\label{fig::hadComp}
\endFigure

\beginFigure
\includegraphics*[width=.48\textwidth]{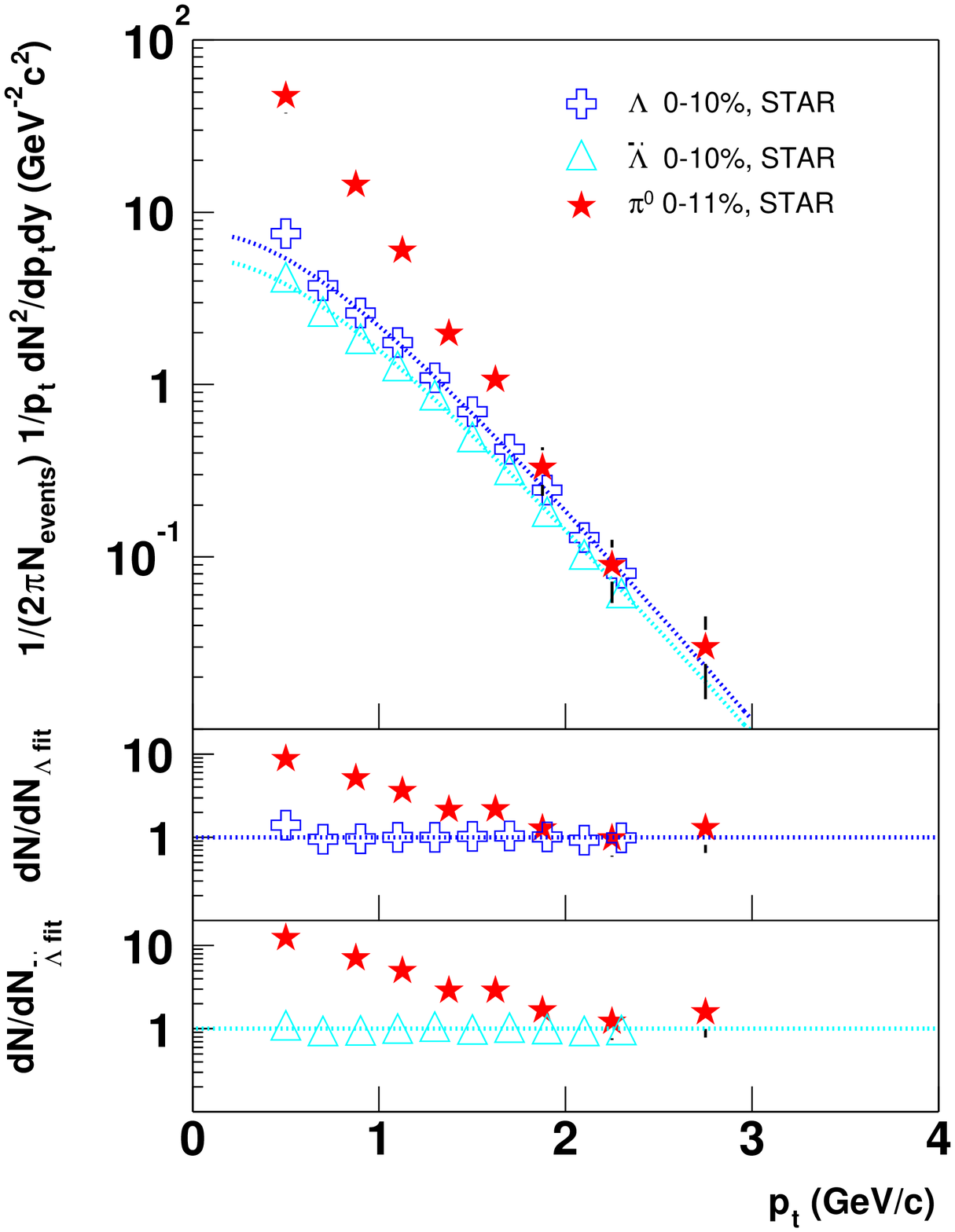}
\caption{Top frame: comparison between STAR {\piz}, and $\Lambda$($\bar{\Lambda}$) measurements \cite{star_lambda} about mid--rapidity for {\auau} collisions at {\sqnn} = 130\gev. Lower frames: ratios of these spectra to Bose--Einstein function fits to the STAR $\Lambda$ and $\bar{\Lambda}$ measurements. For reference, dashed lines indicate $dN/dN_{\Lambda(\bar{\Lambda}) fit}$ of 1. Normalization uncertainties in both the {\piz} ($\pm$49\%) and $\Lambda$($\bar{\Lambda}$) ($\pm$10\%) measurements have not been included with the data points.}
\label{fig::lambdaComp}
\endFigure

A comparison of the {\piz} spectrum for the 0--11\% centrality class was also made to other identified pion spectra for central collisions at {\sqnn} = 130 {\gev}. The PHENIX experiment has published {\piz}, $\pi^+$ and $\pi^-$ spectra for central events \cite{phenix_charged,phenix_suppression}. The {\piz} spectra were measured via the {\pizgg} decay channel using both Lead--Scintillator (PbSc) and Lead--Glass (PbGl) calorimeters. These data overlap the STAR {\piz} measurement in the range 1$<${\pt}$<$3{\gevc}. Ratios between the central STAR {\piz} spectrum to power--law fits of the PHENIX {\piz} spectra indicate that the shapes of the spectra are consistent (Fig. \ref{fig::phenixComp}) although the two experiments have a systematic offset in normalization. In the region of overlap, the STAR {\piz} spectrum is systematically higher than the PHENIX spectra. A systematic difference in the same direction is also observed in comparisons between the charged hadron spectra ($\left(h^+ + h^-\right)/2$) from the two experiments for 1$<$\pt$<$3{\gevc}. Direct comparison of the 0--11\% STAR {\piz} spectrum and the 0--5\% PHENIX $\pi^\pm$ spectrum shows the two are consistent in shape (Fig. \ref{fig::phenixChargedPiComp}), although once again the normalizations are systematically different once a linear scale factor (0.91 $\pm$ 0.04, deduced from $C$ values given in \cite{star_highpt}) is applied to convert the 0--5\% $\pi^\pm$ data to the 0--10\% centrality class. These ratios indicate that meson to baryon ratios are {\it internally} consistent within PHENIX and STAR, although between PHENIX and STAR the normalization of the spectra is systematically shifted for {\pt} between 1{\gevc} and 3\gevc.

\beginFigure
\includegraphics*[width=.48\textwidth]{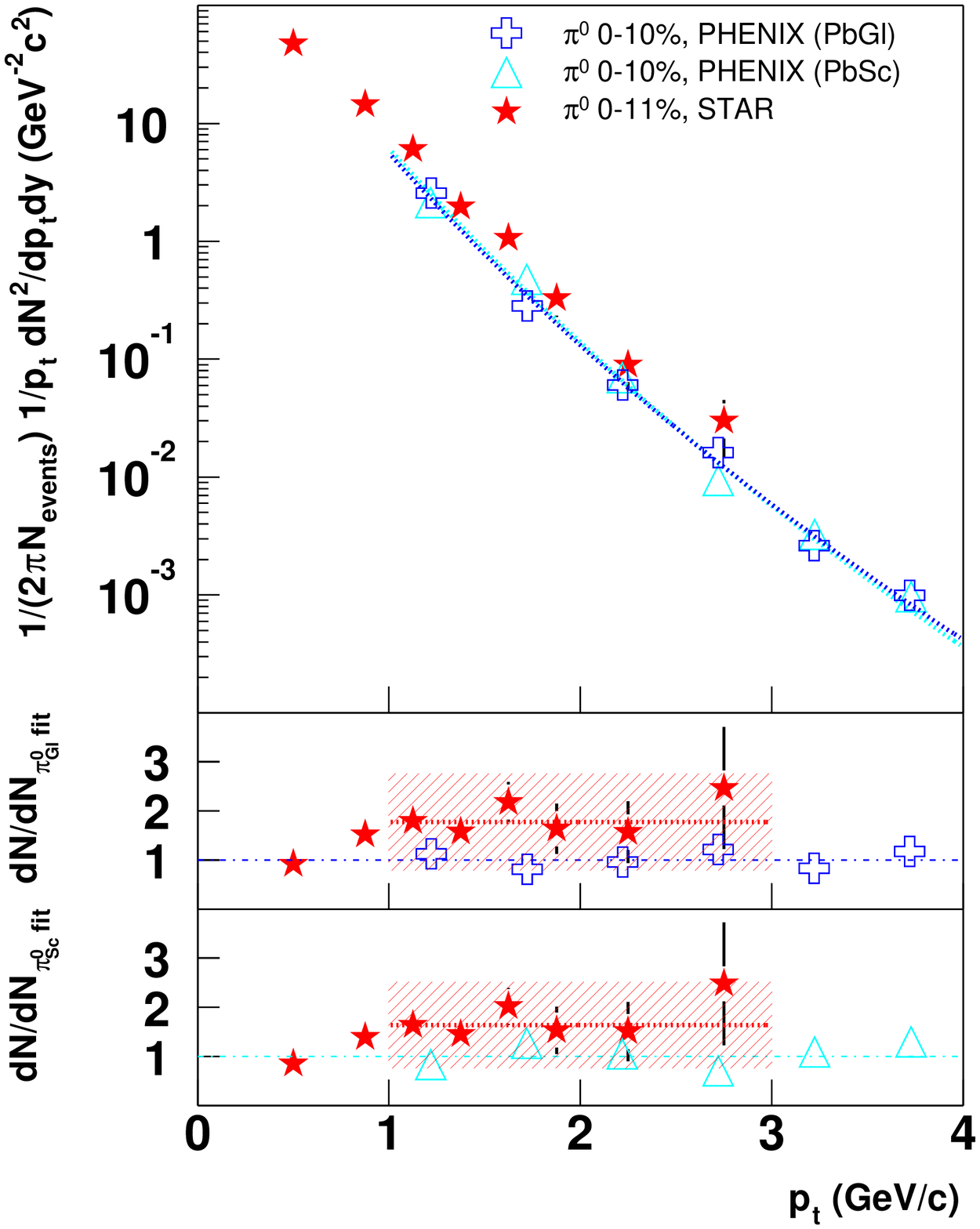}
\caption{Top frame: comparison between the STAR {\piz} measurement and PHENIX {\piz} measurements \cite{phenix_suppression} about mid--rapidity for {\auau} collisions at {\sqnn} = 130\gev. Lower frames: ratios of these spectra to a power law fit to the PHENIX {\piz} measurements. For reference, dashed lines in the lower frames indicate $dN/dN_{PHENIX~\pi^0 fit}$ of 1, and 1.78 $\pm$ 0.98 and 1.64 $\pm$ 0.86 in the middle and bottom frames respectively. Normalization uncertainties in both the STAR ($\pm$49\%) and PHENIX ($\pm$25\% for PbGl and $\pm$20\% for PbSc) {\piz} measurements have not been included with the data points.}
\label{fig::phenixComp}
\endFigure

\beginFigure
\includegraphics*[width=.48\textwidth]{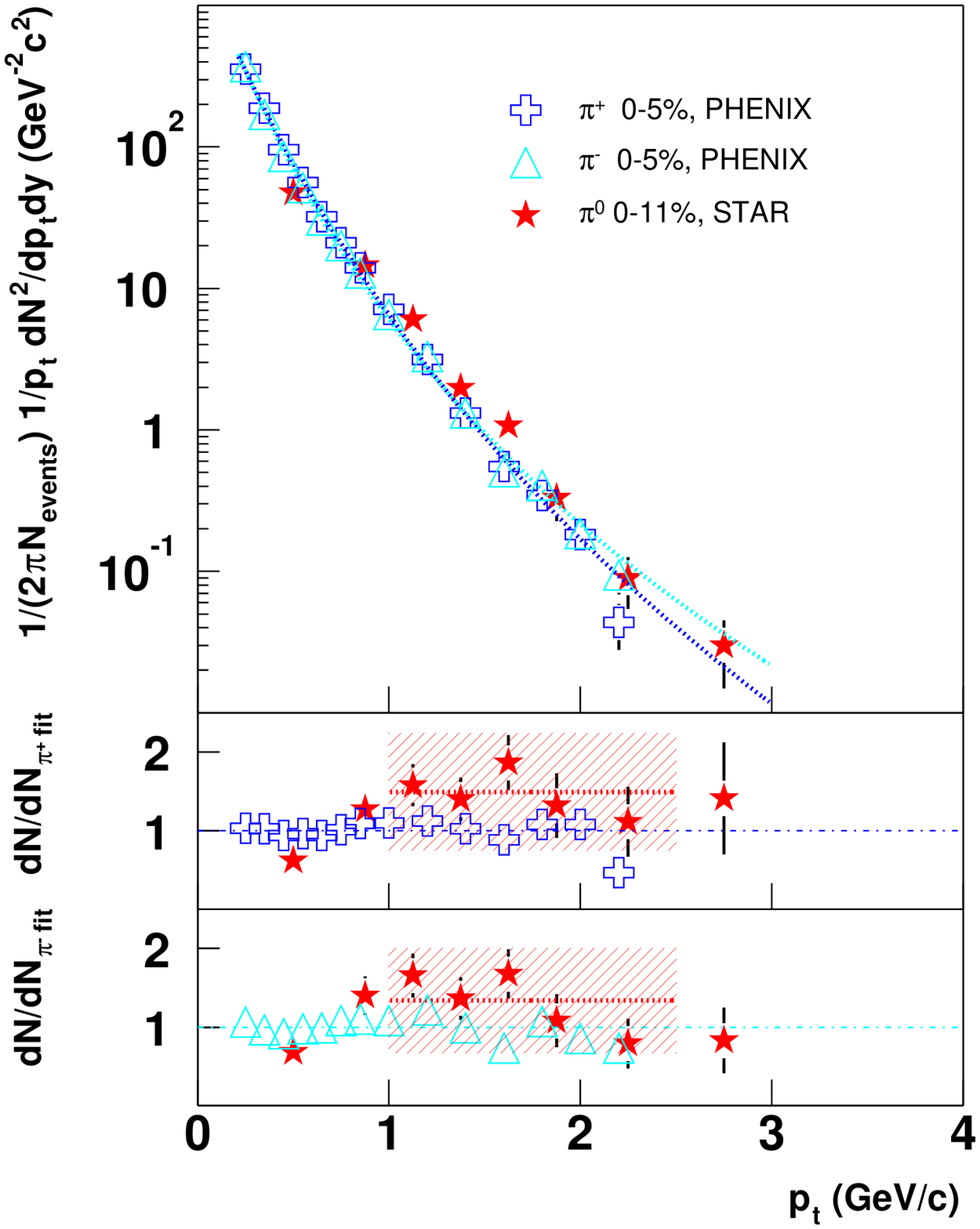}
\caption{Top frame: comparisons between the STAR {\piz} measurement and PHENIX charged pion measurements \cite{phenix_charged} about mid--rapidity for {\auau} collisions at {\sqnn} = 130\gev. Lower frames: ratios of these spectra to a power law fit to the PHENIX charged pion measurements. For reference, dashed lines in the lower frames indicate $dN/dN_{\pi^\pm fit}$ of 1, and 1.49 $\pm$ 0.75 and 1.34 $\pm$ 0.67 in the middle and bottom frames respectively. Normalization uncertainties in both the {\piz} ($\pm$49\%) and charged pion ($\pm$11\%) measurements have not been included with the data points.}
\label{fig::phenixChargedPiComp}
\endFigure

\section{Conclusion}
We have presented the first inclusive mid--rapidity, $|$\y$|$$<$0.5, photon spectra as a function of centrality from {\auau} collisions at {\sqnn} = 130 {\gev}. The spectra of {\piz}s about mid--rapidity, $|$\y$|$$<$1.0, have been presented; as well as the contribution from {\pizgg} decays to the inclusive photon spectrum. Near {\pt} = 1.65{\gevc} the fractional contribution from {\pizgg} decays to the inclusive photon spectrum for the 0--11\% most central collisions begins to decrease significantly. This decrease indicates that relative to the {\pizgg} decay, the contribution from other sources of photons increases with {\pt}. In order to understand the origin of this decrease other electromagnetic decays must be measured or estimated. The combination of increased event statistics in future measurements with the excellent energy resolution achieved using this photon detection technique ($\Delta$E/E $\approx$ 2\% at 0.5\gevc) will make the measurement of the $\eta$ feasible. A statistically significant enhancement in the two--photon invariant mass distribution has already been observed in the vicinity of the $\eta$ mass. Increased event statistics will also lead to higher precision measurements and extend the {\pt} range of the {\piz} spectra. Advances in these directions will not only enhance our understanding of contributions to the single photon spectra, but will also aid measurements of the relative abundance of mesons and baryons at high {\pt} ($>$3{\gevc}) where the expected effects of collective motion become less dominant.

{\bf Acknowledgements:}
We thank the RHIC Operations Group and RCF at BNL, and the NERSC Center at LBNL for their support. This work was supported in part by the HENP Divisions of the Office of Science of the U.S. DOE; the U.S. NSF; the BMBF of Germany; IN2P3, RA, RPL, and EMN of France; EPSRC of the United Kingdom; FAPESP of Brazil; the Russian Ministry of Science and Technology; the Ministry of Education and the NNSFC of China; Grant Agency of the Czech Republic, FOM and UU of the Netherlands, DAE, DST, and CSIR of the Government of India; the Swiss NSF.





\end{document}